\documentclass[preprint,12pt,sort&compress]{elsarticle}

\usepackage{amsmath, amssymb, amsthm}

\usepackage{bm}
\usepackage{graphicx}
\usepackage{graphics}
\usepackage{subfigure}
\usepackage{subcaption}
\usepackage{tabularx,caption}
\usepackage{float}
\usepackage{color}
\usepackage{natbib}
\usepackage{multirow}
\usepackage{booktabs}
\usepackage{diagbox}
\usepackage{algorithm,algpseudocode}
\usepackage{enumitem}
\usepackage{mathtools}   
\AtBeginDocument{%
  \hfuzz=120pt%
  \vfuzz=120pt%
  \hbadness=10000%
  \vbadness=10000%
}

\usepackage{lineno}

\usepackage{graphicx}

\usepackage{makecell}

\journal{}

\begin{document}
\hfuzz=10000pt
\vfuzz=10000pt
\hbadness=10000
\vbadness=10000

\begin{frontmatter}



  \title{Flow-based generative models for amortized Bayesian inference in regression and inverse PDE problems}

  \author[hust]{Shaoqian Zhou}
  \author[shnu]{Ling Guo}
  \author[hust]{Xuhui Meng \fnref{1}}


  \address[hust]{Institute of Interdisciplinary Research for Mathematics and Applied Science,\\ School of Mathematics and Statistics, Huazhong University of Science and Technology, Wuhan 430074, China}
  \address[shnu]{Department of Mathematics, Shanghai Normal University, Shanghai, China}

  \fntext[1]{Corresponding author: xuhui\_meng@hust.edu.cn (Xuhui Meng).}




\begin{abstract}
Bayesian inference provides a principled framework for uncertainty quantification in scientific machine learning. However, conventional Bayesian approaches typically require solving a new inference problem for each observation set, resulting in substantial computational costs that hinder real-time applications such as online monitoring, digital twins, and decision-making under uncertainty. Furthermore, many scientific machine learning problems involve inference over infinite-dimensional function spaces with varying numbers and locations of observations, posing significant challenges for existing amortized inference methods. In this work, we propose Flow-ABI, a flow-based generative framework for amortized Bayesian inference in regression and inverse partial differential equation (PDE) problems. The proposed framework consists of two components: (i) a functional prior model that learns expressive prior in function spaces directly from historical data and physical knowledge through flow matching, and (ii) a set-conditioned functional posterior sampler that performs amortized Bayesian inference by mapping observation sets to posterior distributions over functions. The learned posterior model naturally accommodates varying numbers and locations of observations, is invariant to permutations of the measurement set, and generalizes across different observation discretizations. Once trained, Flow-ABI enables near-real-time posterior sampling for previously unseen observations without retraining or iterative optimization. The proposed methodology can be seamlessly integrated with a broad class of scientific machine learning frameworks, including physics-informed neural networks and neural operators, for uncertainty-aware inverse PDE modeling. Numerical experiments on representative regression and inverse PDE benchmarks demonstrate that Flow-ABI accurately captures both Gaussian and non-Gaussian posterior distributions while achieving speedups of more than two orders of magnitude relative to the gold-standard Bayesian inference method, Hamiltonian Monte Carlo. These results demonstrate that the Flow-ABI as an effective, scalable, and computationally efficient framework for uncertainty quantification in scientific machine learning.
\end{abstract}




\end{frontmatter}


\section{Introduction}
\label{sec:induction}

Deep learning has emerged as a powerful paradigm for scientific computing, achieving remarkable success in a wide range of applications, including function approximation, forward and inverse partial differential equation (PDE) solving \cite{raissi2019physics,sirignano2018dgm}, and operator learning \cite{lu2021learning,li2020fourier,kovachki2023neural}. Despite these advances, observational data in real-world scientific and engineering systems are often sparse, noisy, and incomplete, resulting in significant uncertainty in model predictions \cite{psaros2023uncertainty,zou2024neuraluq,meng2021multi,linka2022bayesian,yang2019adversarial,mollaali2025conformalized}. Quantifying and propagating such uncertainties are essential for reliable decision-making in safety-critical applications, including medical diagnosis, structural health monitoring, climate forecasting, and digital twins.

Bayesian inference provides a principled framework for uncertainty quantification (UQ) by representing unknown quantities as probability distributions and updating prior beliefs using observational data. In the context of deep learning, Bayesian neural networks (BNNs)\cite{neal2012bayesian,pearce2020expressive}, deep ensembles \cite{lakshminarayanan2017simple}, and Monte Carlo dropout \cite{gal2016dropout} have become widely used approaches for uncertainty estimation and have been successfully extended to scientific machine learning (SciML). Examples include Bayesian physics-informed neural networks for solving forward and inverse PDEs \cite{yang2021b} and ensemble-based approaches for uncertainty-aware operator learning \cite{pickering2022discovering}. A central component of Bayesian inference is the prior distribution. For BNNs, it is well known that infinitely wide networks converge to Gaussian processes under suitable assumptions \cite{neal2012bayesian,pearce2020expressive}, while carefully designed parameter priors can induce analytically tractable function-space priors \cite{neal2012bayesian,pearce2020expressive}. More recently, deep generative models such as generative adversarial networks  have been employed to learn expressive functional priors directly from historical data, enabling the representation of complex Gaussian and non-Gaussian stochastic processes \cite{yang2020physics,meng2022learning,zou2023hydra,zhou2025scalable}.

Given a prior and observational data, Bayesian inference proceeds by constructing the posterior distribution. To this end, numerous inference algorithms have been developed, including Markov chain Monte Carlo (MCMC), Hamiltonian Monte Carlo (HMC), variational inference (VI), and Laplace approximations. Comprehensive reviews of uncertainty quantification methods in SciML can be found in \cite{psaros2023uncertainty,zou2024neuraluq}. Despite substantial progress, most existing approaches share a common limitation: posterior inference must be performed from scratch for each new set of observations. Consequently, the computational cost scales with the number of inference tasks, making these methods impractical for applications requiring rapid updates, online monitoring, real-time decision-making, or digital-twin deployment.

Simulation-based inference (SBI), also known as likelihood-free inference \cite{wildberger2023flow,cranmer2020frontier,sharrock2022sequential,geffner2022score,radev2020bayesflow,yuan2026bayesflow,kumar2025bayesflow++}, offers a promising alternative. Instead of solving a new Bayesian inference problem for each observation set, SBI learns a global mapping from observations to posterior distributions during an offline training stage, thereby amortizing the cost of inference across future tasks. Once trained, the resulting model can generate posterior samples for previously unseen observations with negligible additional computational cost. SBI has achieved considerable success in fields such as gravitational-wave inference \cite{dax2021real}, cosmology \cite{leclercq2018bayesian}, and structural health monitoring \cite{ni2022bayesian}.

However, existing SBI methods are not directly applicable to many scientific machine learning problems. First, conventional SBI typically focuses on finite-dimensional parameter vectors, whereas quantities of interest in SciML often reside in infinite-dimensional function spaces, such as spatially varying material properties, source terms, boundary conditions, or PDE solutions. Second, observations in conventional SBI are commonly represented by fixed-dimensional vectors or handcrafted summary statistics. In contrast, observations in SciML typically consist of measurements collected at spatial and temporal locations that may differ in different tasks, and may also form an unordered set. Consequently, the posterior inference operator should produce identical posterior distributions regardless of the ordering of the measurements while accommodating varying numbers and locations of observations. Third, scientific systems are governed by physical laws, usually expressed as PDEs, which impose strong structural constraints on admissible posterior distributions. These characteristics fundamentally distinguish SciML from traditional SBI settings and motivate the development of amortized Bayesian inference methods tailored to function-space learning and physics-constrained systems.

To the best of our knowledge, the only amortized Bayesian inference framework specifically developed for PDE-governed inverse problems is based on the likelihood-free  method in \cite{zeng2025solving}. This approach combines conditional normalizing flows with a convolutional summary network to estimate posterior distributions over high-dimensional groundwater conductivity fields. While demonstrating promising performance, several limitations remain. First, the prior is defined over a finite-dimensional parameter vector, while the induced prior in the underlying function space remains unclear. Second, the use of normalizing flows imposes architectural constraints such as invertibility and tractable Jacobian determinant computation, which can complicate optimization and limit scalability in high-dimensional problems. Third, the summary network is trained with clean data  from numerical simulations, potentially introducing a mismatch between training and deployment conditions. Finally, the convolutional architecture assumes a prescribed ordering of observations and therefore does not naturally guarantee that the inferred posterior remains unchanged under permutations of the measurement set. Moreover, it is not inherently invariant to changes in sensor layouts or observation discretizations.

Motivated by these challenges, we propose a novel framework based on conditional flow matching \cite{wildberger2023flow,lipman2022flow} for amortized Bayesian inference in regression and inverse PDE problems. Unlike normalizing flows, flow matching learns probability transport through vector-field regression and avoids the need for invertible architectures and Jacobian determinant evaluation. Building upon this formulation, we develop a set-conditioned posterior inference framework that learns a mapping from observation sets to posterior distributions over functions. The resulting posterior operator naturally accommodates varying numbers of measurements, remains invariant to permutations of the observation set, and generalizes across different observation discretizations. Furthermore, the framework enables efficient posterior sampling for unseen observations without retraining or iterative optimization, making it particularly attractive for efficient uncertainty quantification.

The main contributions of this work are summarized as follows:
\begin{enumerate}
  \item We develop a one-step flow-matching-based framework for learning expressive function-space priors directly from historical data and physical constraints, avoiding restrictive parametric assumptions on stochastic processes.
 \item We introduce a conditional flow matching approach for amortized Bayesian inference over function spaces. Once trained, the framework enables near-real-time posterior sampling for previously unseen observations.
 \item We put forth a set-conditioned functional posterior model based on conditional flow matching, which learns posterior distributions over functions from sparse and noisy observation sets. The resulting model is permutation-invariant with respect to the observation set and naturally accommodates varying numbers and locations of measurements.
 \item We demonstrate that the proposed framework can be seamlessly integrated with both physics-informed neural networks (PINNs) and neural operators to perform uncertainty-aware inverse PDE modeling. The resulting approach enables efficient Bayesian inversion and uncertainty quantification for PDE-governed systems from sparse and noisy observations.
\end{enumerate}


The rest of this paper is organized as follows: In Sec. \ref{sec:method}, we present the problem formulation as well as the Flow-ABI, i.e. flow-based generative models for amortized Bayesian inference, in regression and inverse differential equations; the numerical results are shown in Sec. \ref{sec:results}, and a summary on this study is present in Sec. \ref{sec:summary}.

\section{Methodology}
\label{sec:method}

\subsection{Problem Formulation}

In this study, we consider two particular scenarios, i.e., function approximation or regression and inverse PDE problems. In the former, we assume that we have measurements collected from sensors on a target function that may be sparse and noisy, and we aim to fit it with uncertainties. As for the latter,  consider a parametric PDE governing the dynamics of a physical system:
\begin{subequations}\label{eq:pde}
  \begin{align}
    &\mathcal{L}_{\bm{\beta}}\bigl[u(t,\bm{x})\bigr]
    = f(t,\bm{x}),
    \quad \bm{x}\in\Omega,\; t\in\Omega_t,\; \bm{\beta}\in B,
    \label{eq:pde_eq}\\
    &\mathcal{B}_{\bm{\beta}}\bigl[u(t,\bm{x}_{\mathrm{bc}})\bigr]
    = b(t,\bm{x}_{\mathrm{bc}}),
    \quad \bm{x}_{\mathrm{bc}}\in\Gamma,
    \label{eq:pde_bc}\\
    &u(0,\bm{x})
    = u_0(\bm{x}),
    \label{eq:pde_init}
  \end{align}
\end{subequations}
where $\bm{x}\in\Omega\subset\mathbb{R}^{D_{\bm{x}}}$ denotes the spatial coordinate on a bounded domain $\Omega$ with boundary $\Gamma$, $\bm{x}_{\mathrm{bc}}\in\Gamma$ is the boundary coordinate, $t\in\Omega_t$ is the temporal coordinate, $\bm{\beta}\in B\subset\mathbb{R}^{D_{\bm{\beta}}}$ is a parameter vector with the probability space $B$, $u\in\mathcal{F}:=L^{2}(\Omega)$ is the solution, $\mathcal{L}_{\bm{\beta}}$ is the any differential operator parametrised by $\bm{\beta}$, $f$ is the source term, $\mathcal{B}_{\bm{\beta}}$ is the operator impose on the boundary values, $b$ is the prescribed boundary condition, and $u_0\in\mathcal{F}$ is the initial condition. The particular inverse problem we consider here is formulated as follows: we assume that we have partial observations on $u$, $f$, and possibly $b$, $u_0$ as well as $\bm{\beta}$, and we would like to reconstruct the full fields of $u$, $f$, and possibly $\bm{\beta}$, with uncertainties.  Further, we denote the quantity of interest (QoI) or the target function in the regression problem as $u(t, \bm{x})$ in what follows without confusion.

In the Bayesian framework, the observations associated with QoI, denoted by $\mathcal{X}$, are modeled as
\begin{equation}\label{eq:obs_additive}
  \mathcal{Y}_{\mathrm{obs}}
  =
  \mathcal{M}(\mathcal{X})
  +
  \bm{\eta},
  \qquad
  \bm{\eta}\sim p_{\bm{\eta}},
\end{equation}
where $\mathcal{X}$ denotes the QoI, which may correspond to the solution field $u$, the source term $f$, the parameter field $\bm{\beta}$, or a combination thereof. Here, $\mathcal{M}$ is the observation operator and $p_{\bm{\eta}}$ denotes the probability distribution characterizing the measurement noise. A common choice is the isotropic Gaussian distribution
$
\bm{\eta}
\sim
\mathcal{N}
\left(
\bm{0},
\sigma_{\mathcal{Y}}^2 \bm{I}
\right).
$
Given a prior distribution
$
\mathcal{X}
\sim
P(\mathcal{X}),
$
Bayesian inference aims to estimate the posterior distribution of $\mathcal{X}$ conditioned on the observations $\mathcal{Y}_{\mathrm{obs}}$. According to Bayes' theorem,
\begin{equation}
P(\mathcal{X}\mid \mathcal{Y}_{\mathrm{obs}})
=
\frac{
P(\mathcal{Y}_{\mathrm{obs}}\mid \mathcal{X})
P(\mathcal{X})
}{
P(\mathcal{Y}_{\mathrm{obs}})
}.
\end{equation}
The objective of the present work is to efficiently approximate this posterior distribution and subsequently quantify the uncertainty of the QoI through posterior samples.



\subsection{Flow-ABI: Flow-based generative models for amortized Bayesian inference}
\label{sec:framework}

\begin{figure}[H]
  \centering
  \includegraphics[width=1.\textwidth]{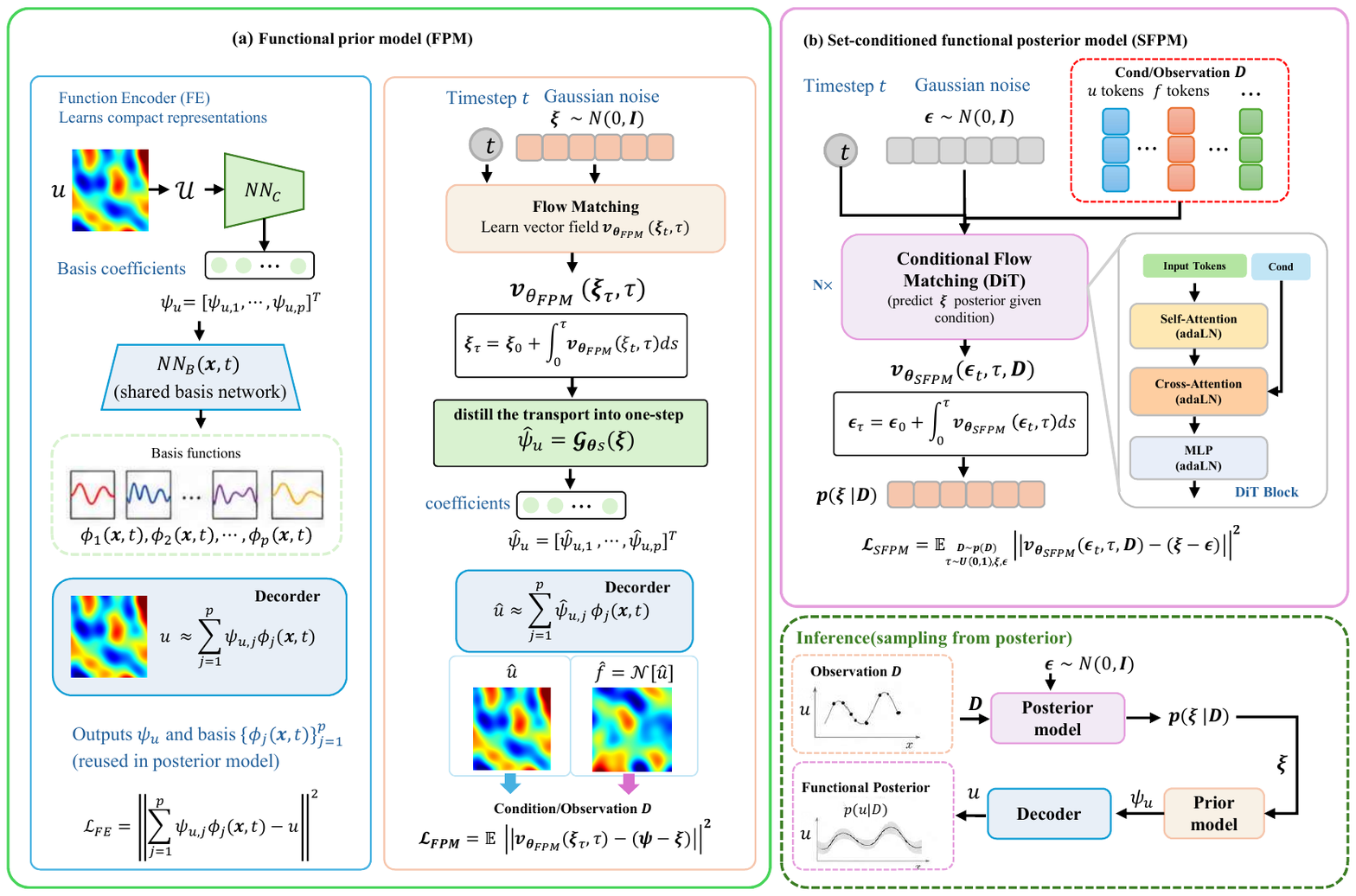}
  \caption{\label{fig:model_architecture}
    Schematic of the Flow-ABI (flow-based generative framework for amortized Bayesian inference), consisting of two components: (a) Functional prior model (FPM) for learning the distribution over function representations from FE, and (b) a Set-conditioned functional posterior model (SFPM)  sampler utilizing the conditional flow matching or diffusion Transformer.
  }
\end{figure}

As shown in Fig. \ref{fig:model_architecture}, the proposed Flow-ABI consists of two components:
\begin{enumerate}
\item A \textbf{Functional prior model} (FPM) via one-step flow matching, which consists of a function encoder (FE) that maps functions into a low-dimensional coefficient space, and a \textbf{one-step flow-matching model} that learns an expressive distribution over the coefficients from the FE.
\item A \textbf{set-conditioned functional posterior model} (SFPM) that performs amortized Bayesian inference via conditional flow matching.
\end{enumerate}


The overall inference pipeline can be summarized as
\[
\mathcal{D}{\rm obs}
\longrightarrow
p(\bm{\xi}|\mathcal{D}{\rm obs})
\longrightarrow
p(\bm{\psi}|\mathcal{D}{\rm obs})
\longrightarrow
p(\mathcal{X}|\mathcal{D}{\rm obs}),
\]
where $\bm{\xi}$ denotes a latent Gaussian variable and $\bm{\psi}$ denotes the latent coefficient representation of the function.

\subsection{Learning functional prior via generative model}
\label{sec:functional_posterior}
We assume that we have historical data for $\mathcal{X}$,  which are either collected from sensors/numerical simulations or realizations of a prescribed functional prior, e.g., a certain Gaussian process (GP). In addition,  we assume each realization is observed at certain number of temporal-spatial locations, yielding a discrete representation for each realization.  We aim to learn the prior represented by the historical data in the functional space using the proposed model.

We now introduce the architectures and training of the proposed model. 
As shown in Fig. \ref{fig:model_architecture}, the functional prior model is composed of two submodels, i.e., a function encoder (FE) and a flow-matching generative model for learning the distributions over the compact representation from FE.  Similar as in \cite{zhou2025scalable}, the FE consists of two subnetworks: a coefficient network $\mathcal{NN}_{C;\bm{\theta}_C}$ and a basis network $\mathcal{NN}_{B;\bm{\theta}_B}$, where $\bm{\theta}_C$ and $\bm{\theta}_B$ are parameters in these two neural networks, respectively.  Without loss of generality, we assume the QoI is $u$. The coefficient network maps the representation $\mathcal{U}$ corresponding to a realization of $u$ to a latent coefficient vector
\begin{equation}
\bm{\psi}(\mathcal{U}_i)
= \mathcal{NN}_{C; \bm{\theta}_C}(\mathcal{U}_i) = 
(\psi_1,\ldots,\psi_p)^T
\in
\mathbb{R}^{p},
\end{equation}
where $\mathcal{U}_i$ denotes the representation for the $i$th realization of $u$, i.e., $u_i = \{u(t_k, \bm{x}_k)\}^{N_{u_i}}_{k=1}$, with $k$ denoting the $k$th measurement of $u_i$, $t_k/\bm{x}_k$ the coordinate, and $N_{u_i}$ the number of measurements for $u_i$. In addition, the basis network learns a set of coordinate-dependent basis functions
\begin{equation}
\bm{\phi}(t, \bm{x})
= \mathcal{NN}_{B;\bm{\theta}_B}(t, \bm{x}) =
\bigl(
\phi_1(t, \bm{x}),
\ldots,
\phi_p(t, \bm{x})
\bigr)^T.
\end{equation}
The target function is then reconstructed through the low-rank expansion
\begin{equation}
u_{\mathrm{FE}}(t, \bm{x}; \mathcal{U}_i)
=
\sum_{j=1}^{p}
\psi_j (\mathcal{U}_i) \phi_j(t,\bm{x}).
\label{eq:function_reconstruction}
\end{equation}
All the parameters in $\mathcal{NN}_{C; \bm{\theta}_C}$ and $\mathcal{NN}_{B; \bm{\theta}_B}$ are trained based on the mean square errors between the $u_{\mathrm{FE}}(t, \bm{x}; \mathcal{U}_i)$ and $u_i$. This representation projects the original infinite-dimensional function onto a finite-dimensional coefficient space while preserving its continuous dependence on the spatial coordinate.  Specifically, the representation for the realization of $u_i$ can be obtained using either the discrete function values as in DeepONet \cite{lu2021learning}, or the leading eigenvalues computed from the principal component analysis (PCA) as \cite{zhou2025scalable}. The resulting feature vector $\bm{\psi}(\mathcal{U})$ serves as a compact descriptor of the each realization from the stochastic process represented by the historical data.


Having obtained the latent coefficient representations, we next learn their empirical distribution. Let
$
p_{\bm{\psi}}
$
denote the distribution of the latent coefficients extracted by the FE. To approximate this distribution, we employ a two-stage flow-matching framework consisting of a base flow model and a distilled one-step generator. In the first stage, a flow-matching model learns a transport from a standard Gaussian base distribution
$
p_0(\bm{\xi})
=
\mathcal{N}(\bm{0},\bm{I}),
$
to the target coefficient distribution
$
p_1(\bm{\psi})
\approx
p_{\bm{\psi}}.
$
The transport is parameterized by a velocity field
\[
v_{\bm{\theta}_{\mathrm{FPM}}}:
\mathbb{R}^{p}\times [0,1]
\rightarrow
\mathbb{R}^{p},
\]
through the ordinary differential equation
\begin{equation}
\frac{\mathrm{d}\bm{\xi}_{\tau}}
{\mathrm{d}\tau}
=
v_{\bm{\theta}_{\mathrm{FPM}}}
(
\bm{\xi}_{\tau},
\tau
),
\qquad
\tau\in[0,1].
\end{equation}
Following the conditional flow-matching formulation, a linear probability path is constructed between a Gaussian sample
$
\bm{\xi}
\sim
\mathcal{N}(\bm{0},\bm{I})
$
and a target coefficient sample
$
\bm{\psi}
\sim
p_{\bm{\psi}},
$
according to
\begin{equation}
\bm{\xi}_{\tau}
=
(1-\tau)\bm{\xi}
+
\tau\bm{\psi}.
\end{equation}
The corresponding target vector field is
\[
\frac{\mathrm{d}\bm{\xi}_{\tau}}
{\mathrm{d}\tau}
=
\bm{\psi}
-
\bm{\xi},
\]
which leads to the flow-matching objective
\begin{equation}
\mathcal{L}_{\mathrm{FPM}}
(
\bm{\theta}
)
=
\mathbb{E}
\left[
\left\|
v_{\bm{\theta}_{\mathrm{FPM}}}
(
\bm{\xi}_{\tau},
\tau
)
-
(
\bm{\psi}
-
\bm{\xi}
)
\right\|_2^2
\right].
\label{eq:prior_fm}
\end{equation}

Although the trained flow model accurately approximates the latent distribution, generating samples requires numerical integration of the ODE. To accelerate sampling, we further distill the learned transport into a one-step generator following \cite{liu2022flow}. Specifically, a student network
\[
\mathcal{G}_{\bm{\theta}_s}:
\mathbb{R}^{p}
\rightarrow
\mathbb{R}^{p}
\]
is trained to directly map Gaussian noise to latent coefficients. The distillation objective is given by
\begin{equation}
\mathcal{L}_{\mathrm{Distill}}
(
\bm{\theta}_s
)
=
\mathbb{E}_{\bm{\xi}\sim\mathcal{N}(\bm{0},\bm{I})}
\left[
\left\|
\mathcal{G}_{\bm{\theta}_s}
(
\bm{\xi}
)
-
\mathrm{ODESolve}
(
v_{\bm{\theta}_{\mathrm{FPM}}},
\bm{\xi},
0,
1
)
\right\|_2^2
\right],
\label{eq:distill}
\end{equation}
where $\mathrm{ODESolve}(\cdot)$ denotes the terminal state generated by the pretrained flow model.

After training, the distilled generator serves as an efficient latent prior model and produces coefficient samples according to
\begin{equation}
\bm{\psi}_{\mathrm{prior}}
=
\mathcal{G}_{\bm{\theta}_s}
(
\bm{\xi}
),
\qquad
\bm{\xi}
\sim
\mathcal{N}
(
\bm{0},
\bm{I}
).
\label{eq:latent_prior}
\end{equation}

Combining the generated coefficients with the learned basis functions yields a functional prior in the original function space:
\begin{equation}
u_{\mathrm{prior}}
(
t,\bm{x};
\bm{\xi}
)
=
\sum_{i=1}^{p}
\left[
\mathcal{G}_{\bm{\theta}_s}
(
\bm{\xi}
)
\right]_i
\phi_i
(
t,\bm{x}
).
\label{eq:function_prior}
\end{equation}
Consequently, the proposed framework learns a probability measure over functions by modeling the latent coefficient distribution and subsequently mapping it back to the function space through the learned basis expansion.

For inverse PDE problems governed by Eq.~\eqref{eq:pde}, we assume that either historical realizations of the target field or samples from a prescribed prior distribution are available. Following the procedure described above, these realizations are first used to train the functional prior model and thereby learn a functional prior distribution for the QoI. Once the functional prior has been established, prior realizations of other quantities appearing in Eq.~\eqref{eq:pde}, such as source terms, boundary conditions, or initial conditions, can be generated through the governing PDE using automatic differentiation (AD). Specifically, sampled realizations of $u_{\mathrm{prior}}$ are substituted into Eq.~\eqref{eq:pde}, and the corresponding realizations of $f$, $b$, or $u_0$ are computed via AD, similar to the approaches adopted in \cite{raissi2019physics,meng2022learning,zhou2025scalable,psaros2023uncertainty,zou2024neuraluq,zou2023hydra}.

As an alternative to AD, pretrained neural operators (NOs), such as the deep operator networks (DeepONet) \cite{lu2021learning} or the Fourier Neural Operator (FNO), can be employed as surrogate models to capture the relationships among different quantities in Eq.~\eqref{eq:pde}. Without loss of generality, suppose a pretrained neural operator $\mathcal{N}$ which maps $f$ to $u$, i.e., 
$
u=\mathcal{N}(f)
$
is available. Following \cite{meng2022learning}, the dataset used to train the neural operator may be regarded as samples from the prior distribution of $f$. The functional prior model can therefore be trained using these samples to learn a functional prior distribution for $f$, denoted by $f_{\mathrm{prior}}(t,\bm{x}; \bm{\xi]})$. New realizations of the solution field can then be generated through the pretrained neural operator according to
$
u_{\mathrm{prior}}
=
\mathcal{N}
\bigl(
f_{\mathrm{prior}}
\bigr).
$
Additional details regarding uncertainty quantification for neural operators can be found in \cite{meng2022learning,psaros2023uncertainty,zou2024neuraluq}.

For problems in which the parameter $\bm{\beta}$ in Eq.~\eqref{eq:pde} is an unknown finite-dimensional vector, we assume that historical samples of $\bm{\beta}$ are available. In this case, the parameter vector is concatenated with the latent coefficient vector $\bm{\psi}$, and the joint distribution is learned using the flow-matching model $\mathcal{G}_{\bm{\theta}_s}(\bm{\xi})$ in Eq.~\eqref{eq:latent_prior}, following \cite{yang2020physics,zhou2025scalable}. If $\bm{\beta}$ represents an unknown field rather than a finite-dimensional parameter, an additional FE is introduced to learn a latent coefficient representation and corresponding basis functions for $\bm{\beta}$. The resulting latent coefficients are then concatenated with those associated with $u$ or $f$, and the joint latent distribution is subsequently learned through the flow-matching framework described in Eqs.~\eqref{eq:latent_prior} - \eqref{eq:function_prior}.

\subsection{Amortized Bayesian inference via a set-conditioned functional posterior model}

In the set-conditioned functional posterior model (SFPM) (Fig. \ref{fig:model_architecture}), we train a conditional flow-matching model, more specifically the diffusion Transformer (DiT) \cite{peebles2023scalable}, to map observational sets to posterior distributions over latent variables, for amortized Bayesian inference. In particular, we seek to approximate the posterior distribution
$
q_{\rm post}^{\bm{\xi}}
(\bm{\xi}\mid \mathcal{D}_{\rm obs}),
$
from which posterior distributions over the latent coefficients and functions can be obtained through the deterministic mappings introduced in the previous subsection.

Similarly, we take the case  where the QoI is $u$ as an example to introduce the architecture as well as training of the SFPM. To construct training data for the posterior model, we leverage the pretrained functional prior. Specifically, latent variables are first sampled from the base Gaussian distribution,
$
\bm{\xi}^{(k)}
\sim
\mathcal{N}(\bm{0},\bm{I}),
$
and subsequently mapped to the function space through the learned prior model based on Eqs.~\eqref{eq:latent_prior} - \eqref{eq:function_prior}, which is denoted as $
u^{(k)} = 
\mathcal{D}
\Bigl(
\mathcal{G}_{\bm{\theta}_s}
(\bm{\xi}^{(k)})
\Bigr)
$ for simplicity.  For each generated realization, a corresponding observation set is constructed according to the measurement model
\[
\mathcal{D}^{(k)}_{\rm obs} = 
\{u^{(k)}(t_j,\bm{x}_j)+{\eta}^{(k)}_j\}_{j=1}^{N_{\bm{u}}},
\]
where the number of observations $N_{u}$ are a random integer whitin the range of the length for the squence, and the spatio-temporal coordinates $\{(t_j,\bm{x}_j)\}_{j=1}^{N_{\bm{x}}}$ are randomly generated within the computational domain. In addition, the noise term ${\eta}^{(k)}$ is drawn from a prescribed measurement-noise distribution to emulate measurement errors in practical applications. Note that the observations on $f$ and/or $\bm{\beta}$ can either be obtained in the same way as generating observations on $u$, or derived from the $u^{(k)}$ using AD or pretrained NO as discussed in the previous subsection.  This procedure yields paired samples
$
(
\bm{\xi}^{(k)},
\mathcal{D}^{(k)}{\rm obs}
)_{k=1}^{N},
$
which can be regarded as samples from the joint distribution
\begin{equation}
(\bm{\xi},\mathcal{D}{\rm obs})
\sim
p(\bm{\xi},\mathcal{D}{\rm obs}).
\label{eq:joint_distribution}
\end{equation}
These paired samples are subsequently used to train the SFPM.

As shown in Fig. \ref{fig:model_architecture}, the SFPM receives three inputs: (i) the noised latent variable $\bm{\epsilon}_{\tau} \in \mathbb{R}^{p}$ at diffusion time $\tau$, (ii) the scalar timestep $\tau \in [0,1]$, and (iii) the observation set $\mathcal{D}_{\rm obs} = \{(t_i, \bm{x}_i, \mathcal{X}_i, c_i)\}_{i=1}^{N^{\mathcal{X}}_{\rm obs}}$ as  condition, where $\mathcal{X}_i$ is the observed value at coordinate $(t_i, \bm{x}_i)$ and $c_i$ is a categorical indicator distinguishing different measurement types (e.g., observations of $u$ versus the forcing term $f$).  Specifically, cross-attention layers are utilized in the SFPM, with the noised latent $\bm{\epsilon}_{\tau}$ as queries, and the observation tokens as keys and values.

To enable the discretization invariance of the SFPM, each observation is tokenized into an embedding vector by encoding its physical coordinate via random Fourier features (RFF) \cite{rahimi2007random}, and then concatenating the resulting coordinate embedding with the observed value and type indicator. Specifically, the temporal-spatial coordinate $(t_i, \bm{x}_i)$ is first lifted into a high-dimensional feature space through the RFF mapping
\begin{equation}
\gamma(t_i, \bm{x}_i)
=
\Bigl[
\cos(\bm{\omega}_1^T \bm{z}_i),\,
\sin(\bm{\omega}_1^T \bm{z}_i),\,
\ldots,\,
\cos(\bm{\omega}_{D/2}^T \bm{z}_i),\,
\sin(\bm{\omega}_{D/2}^T \bm{z}_i)
\Bigr]^T
\in \mathbb{R}^{D},
\label{eq:rff}
\end{equation}
where $\bm{z}_i = (t_i, \bm{x}_i^T)^T$ is the concatenated coordinate vector 
and $\{\bm{\omega}_d\}_{d=1}^{D/2}$ are random frequency vectors sampled from a 
prescribed spectral distribution, e.g., 
$\bm{\omega}_d \sim \mathcal{N}(\bm{0}, \sigma^2 \bm{I})$. The RFF mapping approximates a shift-invariant kernel $k(\bm{z}_i, \bm{z}_j) \approx \gamma(\bm{z}_i)^T \gamma(\bm{z}_j)$ \cite{rahimi2007random}, thereby encoding the {\emph{relative}} positional relationship between any two observation locations through their inner product in the feature space, while keeping each token's coordinate representation absolute and order-independent. The full observation token is then formed as
\begin{equation}
\bm{e}_i 
= 
\mathrm{Linear}
\Bigl(
\bigl[
\gamma(t_i, \bm{x}_i);\, \mathcal{X}_i;\, c_i
\bigr]
\Bigr)
\in \mathbb{R}^{d_{\rm model}},
\label{eq:token_embed}
\end{equation}
where $[\,;\,]$ denotes concatenation, and $\mathrm{Linear}(\cdot)$ projects the concatenated vector to the model dimension $d_{\rm model}$.

Crucially, because each token $\bm{e}_i$ already encodes the absolute 
temporal-spatial coordinate $(t_i, \bm{x}_i)$ of the measurement as part 
of its content, no additional positional encoding is required or applied to 
the observation tokens. As a result, the set of observation embeddings 
$\{\bm{e}_i\}_{i=1}^{N_{\rm obs}}$ is treated as an \emph{unordered} set: 
permuting the observations leaves every token embedding unchanged, so the 
cross-attention output—and hence the predicted vector field—is invariant to 
the ordering of $\mathcal{D}_{\rm obs}$. This design enforces the 
permutation-invariance property required by Bayesian inference, namely
\begin{equation}
q_{\rm post}
\bigl(
u \mid \mathcal{D}_{\rm obs}
\bigr)
=
q_{\rm post}
\bigl(
u \mid \pi(\mathcal{D}_{\rm obs})
\bigr),
\qquad \forall\, \pi \in S_{N_{\rm obs}},
\label{eq:permutation}
\end{equation}
where $S_{N_{\rm obs}}$ denotes the permutation group of $N_{\rm obs}$ elements.

Furthermore, attention masking is applied to the cross-attention layers 
in which the noised latent $\bm{\epsilon}_{\tau}$ (as queries) attends 
to the observation tokens (as keys and values). Specifically, a binary 
mask $\bm{M} \in \{0,1\}^{N_{\rm obs}}$ is constructed for each sample in 
a training batch, where $M_i = 0$ marks a padded or absent observation token 
that is excluded from the attention computation. This mechanism serves two 
purposes: (i) it allows the model to handle observation sets of \emph{varying 
cardinality} $N^{\mathcal{X}}_{\rm obs}$ within the same batch by padding shorter sequences to a fixed length and masking the padded positions, and (ii) it naturally 
supports \emph{partial observations}, where only a subset of physical 
quantities is measured at any given location, by masking the corresponding 
absent tokens. Consequently, the conditioning signal seen by each latent 
query token is computed exclusively from valid, unmasked observations, 
ensuring that the inferred posterior is determined solely by the available 
data rather than by padding artifacts.

We now discuss the training of the SFPM. Let
$
\bm{\epsilon}
\sim
\mathcal{N}
(\bm{0},\bm{I})
$
be a base Gaussian variable and define the probability path
\[
\bm{\epsilon}_{\tau}
 = 
(1-\tau)\bm{\epsilon}
+
\tau\bm{\xi},
\qquad
\tau\in[0,1].
\]
The SFPM parameterizes a conditional vector field
$
v_{\bm{\theta}_{\rm{SFPM}}}
(
\bm{\epsilon}_{\tau},
\tau,
\mathcal{D}_{\rm obs}
;
\bm{\theta}_{\rm DiT}
),
$
which is trained using the flow-matching objective
\begin{equation}
\mathcal{L}_{\rm SFPM}
= 
\mathbb{E}_{\tau \sim \mathcal{U}(0,1), \bm{\epsilon}, \bm{\xi},\mathcal{D}_{\rm obs}}
\left[
\left\|
v_{\bm{\theta}_{\rm{SFPM}}}
(
\bm{\epsilon}_{\tau},
\tau,
\mathcal{D}_{\rm obs}
;
\bm{\theta}_{\rm DiT}
)
- 
(
\bm{\xi}
 -
\bm{\epsilon}
)
\right\|_2^2
\right].
\label{eq:fm_loss}
\end{equation}
After training, posterior samples are generated by integrating the learned conditional flow,
\begin{equation}
\bm{\xi}_{\rm post}
=
\bm{\epsilon}
+
\int_0^1
v_{\bm{\theta}_{\rm{SFPM}}}
(
\bm{\epsilon}_{\tau},
\tau,
\mathcal{D}_{\rm obs}
;
\bm{\theta}_{\rm DiT}
)d\tau.
\label{eq:posterior_latent}
\end{equation}
The resulting posterior coefficient samples are obtained through
\[
\bm{\psi}_{\rm post}
=
\mathcal{G}_{\bm{\theta}_s}
(
\bm{\xi}_{\rm post}
),
\]
and posterior function samples are reconstructed as
\begin{equation}
u_{\rm post}(\bm{x})
=
\sum_{i=1}^{p}
[\bm{\psi}_{\rm post}]_i
\phi_i(t,\bm{x}).
\label{eq:posterior_function}
\end{equation}
Repeating this procedure yields samples from the approximate posterior distribution
\[
\hat{Q}_{\rm post}
(
u
\mid
\mathcal{D}_{\rm obs}
).
\]

After obtaining posterior samples of the primary quantity of interest, posterior distributions of other physical variables appearing in Eq.~\eqref{eq:pde} can be readily derived. For example, if the posterior samples correspond to the solution field $u_{\rm post}$, posterior realizations of the source term $f$, boundary conditions $b$, or other PDE quantities can be computed by substituting $u_{\rm post}$ into the governing equation and evaluating the required differential operators through AD. Alternatively, when a pretrained neural operator (NO) is available to represent the relationship among physical quantities, uncertainty can be propagated through the neural operator. For instance, if posterior samples of the source term are inferred, i.e.,
\[
f_{\rm post}
\sim
\hat{Q}_{\rm post}
(
f
\mid
\mathcal{D}_{\rm obs}
),
\]
the corresponding posterior samples of the solution field can be obtained through
\[
u_{\rm post}
=
\mathcal{N}
(
f_{\rm post}
).
\]
In this manner, the proposed framework naturally propagates uncertainties across coupled physical quantities without requiring additional posterior samplers.

Repeating the above procedure yields samples from the approximate posterior distribution
\[
\hat{Q}_{\rm post}
(
u
\mid
\mathcal{D}_{\rm obs}
),
\]
from which posterior statistics such as the mean, variance, credible intervals, and other uncertainty measures can be readily estimated. Furthermore, since posterior samples are generated independently, the sampling procedure can be fully parallelized through batched computation on modern hardware accelerators, enabling efficient uncertainty quantification for large-scale regression and inverse PDE problems.

Consequently, Flow-ABI learns a permutation-invariant and discretization-invariant set-to-distribution mapping that directly approximates posterior distributions in function space. Unlike conventional inference methods tied to fixed sensor layouts or discretizations, the proposed framework naturally accommodates varying numbers, locations, and types of measurements, enabling scalable amortized Bayesian inference and uncertainty quantification across a broad class of regression and inverse PDE problems.

\section{Results and discussion}
\label{sec:results}
In this section, we will test the effectiveness of Flow-ABI in regression as well as inverse PDE problems with uncertainties. We further showcase the efficiency of the Flow-ABI for posterior estimate using examples of inverse PDE problems.  In addition, we employ the Hamiltonian Monte Carlo (HMC) to estimate the posterior in cases where we are not able to obtain the exact solutions. Specifically, the HMC is implemented based on the No-U-Turn in BlackJAX\cite{cabezas2024blackjax}, which eliminates the need to set the number of steps.  Details on the computations in each test case are present in \ref{sec:computations}. 

\subsection{1D function approximation}
\label{sec:regression}

We first evaluate the proposed Flow-ABI framework on a series of one-dimensional regression problems involving different prior distributions and measurement noise models. Specifically, three representative scenarios are considered: (a) a Gaussian process (GP) prior with a squared exponential kernel and Gaussian measurement noise; (b) a non-Gaussian process prior with Gaussian measurement noise; and (c) a GP prior with a Mat\'{e}rn kernel and Student's t-distributed measurement noise. These cases are designed to assess the ability of the proposed method in cased with both Gaussian and non-Gaussian priors and likelihood models. The corresponding prior specifications and measurement noise models are summarized in Table \ref{tab:1d_regression}.

\begin{table}[ht]
  \centering
  \caption{1D regression problems: Prior and measurement error or noise in each test case.}
  \label{tab:1d_regression}
  \renewcommand{\arraystretch}{1.3}
  \begin{tabularx}{\textwidth}{c >{\centering\arraybackslash}X >{\centering\arraybackslash}X >{\centering\arraybackslash}X}
    \toprule
    & Prior & Kernel & Noise \\
    \midrule
    case (a)
    & $\mathcal{GP}(0, \kappa(x, x'))$
    & Squared exponential, $l = 0.1$
    & $\mathcal{N}(0, 0.1^2)$ \\[4pt]
    case (b)
    & $\exp[\mathcal{GP}(0, \kappa(x, x'))]$
    & Squared exponential, $l = 0.1$
    & $\mathcal{N}(0, 0.1^2)$ \\[4pt]
    case (c)
    & $\mathcal{GP}(0, \kappa(x, x'))$
    & Mat\'{e}rn, $\nu=2.5$, $l = 0.1$
    & $\mathrm{Student's ~t}~(\nu=3,\ \sigma=0.1)$ \\
    \bottomrule
  \end{tabularx}
\end{table}

To begin with, we assess the ability of the functional prior model within Flow-ABI to learn diverse prior distributions in function space from data. For each prior considered in Table~\ref{tab:1d_regression}, we randomly generate 10,000 realizations from the corresponding stochastic process as training samples. Each realization is discretized using 128 uniformly spaced points over the computational domain $x\in\Omega$, where $\Omega = [-1, 1]$. These snapshots are subsequently used to train the function representation module, comprising the coefficient network $\mathcal{NN}_C$ and the basis network $\mathcal{NN}_B$. For this one-dimensional example, the discretized function values at the 128 measurement locations are provided as input to $\mathcal{NN}_C$, while the spatial coordinate $x\in\Omega$ serves as the input to $\mathcal{NN}_B$. The resulting latent coefficients learned by the function encoder are then used to train the one-step flow matching model, which approximates the prior distribution in the latent coefficient space. Samples from the learned functional prior can subsequently be generated according to Eq. \eqref{eq:function_prior}.

\begin{figure}[H]
  \centering
  \includegraphics[width=1\textwidth]{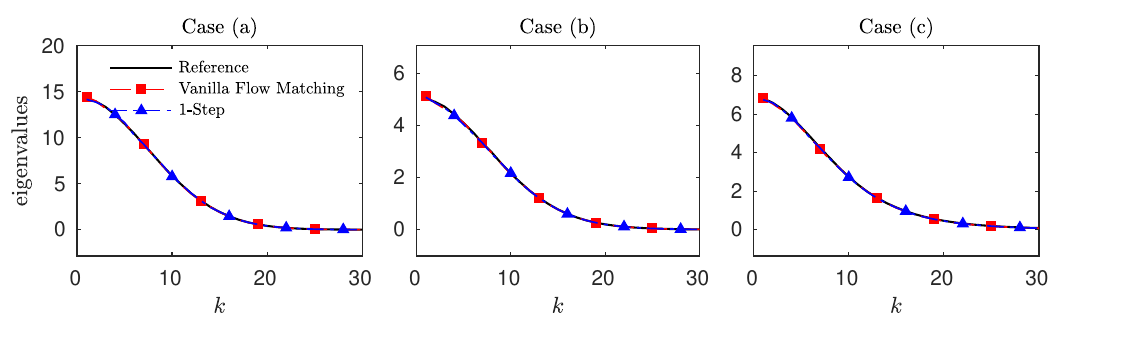}
  \caption{Flow-ABI for 1D regression problem: Eigenvalues of the covariance matrix for the learned functional prior for cases (a)-(c) (from left to right). Black solid line: reference solution, line with squares: preditions from base flow matching model, line with triangles: predictions from one-step generator.}
  \label{fig:eigenvalues_comparison}
\end{figure}

To quantify the accuracy of the learned functional prior, we generate 10,000 samples from the trained model as well as the reference stochastic process, and then we compute the covariance matrices from both sets of samples and perform an eigenvalue decomposition. Fig. \ref{fig:eigenvalues_comparison} presents the leading eigenvalues of the covariance matrices obtained from the reference and learned priors. As can be observed, the proposed functional prior model accurately reproduces the dominant covariance structures of all three stochastic processes, including both Gaussian and non-Gaussian cases. These results demonstrate that the learned prior provides a faithful approximation of the underlying probability measure in function space.

Furthermore, by employing a one-step flow matching distillation strategy, the prior generation process can be performed with a single  network forward evaluation, eliminating the need for iterative numerical integration. This significantly reduces the computational cost of sampling and provides an efficient mechanism for generating large numbers of prior realizations, which is particularly advantageous for constructing training data for the subsequent amortized Bayesian inference stage.

We next train the SFPM using paired training data generated from the pretrained functional prior, as described in Sec. \ref{sec:method}. After training, the posterior model is evaluated on three representative test cases corresponding to scenarios (a)–(c). To assess the generalization capability of the proposed framework, the testing cases involve different target functions as well as observation sets with varying numbers and locations of measurements. In particular, the locations of the measurements in each test case are randomly selected within the range $x \in  [-1, 1]$. These configurations are not tied to any specific sensing pattern and therefore provide a stringent test of the model's ability to perform inference under different observation settings.

For case (a), where both the prior and likelihood are Gaussian, the posterior distribution can be computed analytically using Gaussian process regression (GPR), which is adopted as the reference solution. For cases (b) and (c), where either the prior distribution or the measurement noise model is non-Gaussian, posterior samples are generated using Hamiltonian Monte Carlo (HMC) and serve as the reference solutions. In all cases, the posterior mean and uncertainty are estimated from 10,000 posterior samples with the Monte Carlo method.

\begin{figure}[H]
  \centering
  \includegraphics[width=1\textwidth]{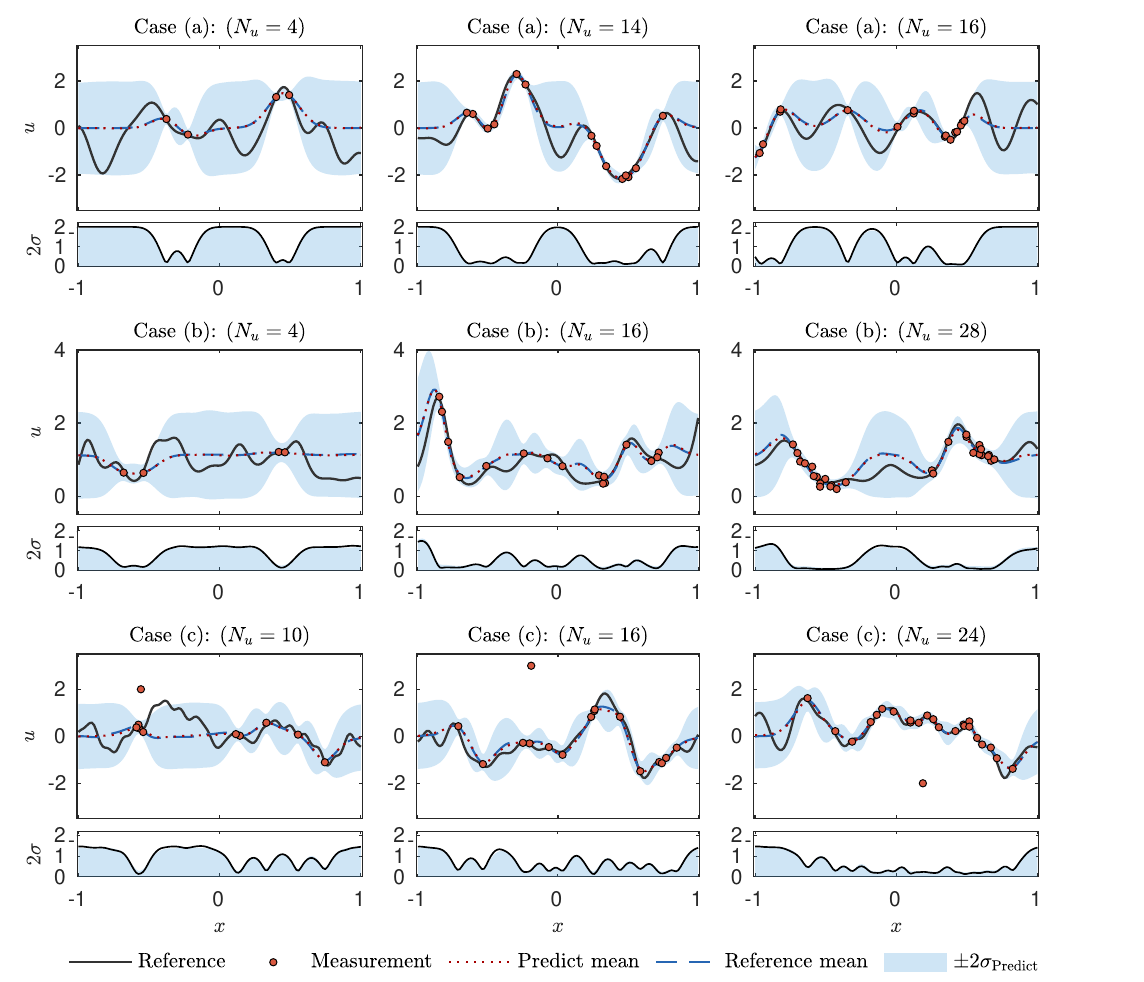}

  \caption{\label{fig:process} Flow-ABI for 1D regression problem. Case (a): Gaussian Process (GP) with length-scale $l = 0.1$ and Gaussian noise $\epsilon_u \sim \mathcal{N}(0,0.1^2)$; Case (b): non-GP method with length-scale $l = 0.1$ and Gaussian noise $\epsilon_u \sim \mathcal{N}(0,0.1^2)$; Case (c): Gaussian Process (GP) with Mat{$\acute{\mbox{e}}$}rn kernel $\nu=2.5$, length-scale $l = 0.1$, and Student-t noise $\epsilon_u \sim \mathrm{StudentT}(\nu=3, \sigma=0.1)$. Black solid: references for the target function and two standard deviations in the uppper and lower parts of each plot, respectively.}
\end{figure}

We illustrate the predictions obtained by Flow-ABI with the corresponding reference solutions in Fig. \ref{fig:process}. Excellent agreement is observed for both the predicted mean and uncertainties across all three scenarios. In particular, the proposed framework accurately captures the posterior distributions arising from both Gaussian and non-Gaussian priors and likelihoods. Moreover, despite the varying numbers and locations of measurements in the testing cases, the prediction accuracy remains consistently high, demonstrating the ability of the proposed approach to effectively process observation sets of varying  size and configuration. These results validate the proposed amortized Bayesian inference framework and indicate that it can accurately approximate complex posterior distributions while retaining the flexibility required for practical scientific machine learning applications involving sparse, noisy, and irregular observations.

To assess the permutation-invariance property of Flow-ABI, we consider a representative case from scenario~(a), where 16 measurements are available. Specifically, we randomly permute the ordering of the observation set while keeping the latent noise realization $\bm{\epsilon}$ fixed, and then generate posterior samples using the trained SFPM. The resulting samples are compared with those obtained from the original observation ordering, as illustrated in Fig.~\ref{fig:process}. Since the latent noise is identical in both cases, any discrepancy can be attributed to the ordering of the observations as well as the ODE integration errors incurred during posterior sample generation.

Let $\bm{\xi}_1,\bm{\xi}_2\in\mathbb{R}^{N\times D}$ denote the two sets of posterior latent samples obtained from the original and permuted observation sets, respectively, where $N=100$ is the number of generated samples and $D=64$ is the latent dimension. We quantify the difference between the two sample sets using the mean absolute error (MAE), defined as
\begin{equation}
E_{\mathrm{MAE}}
=
\frac{1}{ND}
\sum_{i=1}^{N}
\sum_{j=1}^{D}
\left|
\xi_{1,i,j}
-
\xi_{2,i,j}
\right|,
\end{equation}
where $\xi_{1,i,j}$ and $\xi_{2,i,j}$ denote the $j$-th latent component of the $i$-th posterior sample from the original and permuted observation sets, respectively.
For this particular example, the computed MAE is $5.32 \times 10^{-5}$.  This error may be attributed to computational propagation errors within the network. The result confirms that the learned posterior distribution is effectively invariant to permutations of the observation set, thereby verifying that the set-conditioned architecture successfully learns the permutation-invariant posterior operator required for Bayesian inference.

We further demonstrate the capability of Flow-ABI for active learning in regression problems. Specifically, we consider the target function
\[
u(x)=2e^{-(x+1)}\sin(4\pi x)+x^2,
\]
and the prior as well as the noise model are the same as in case (a). We assume that initially we have 17 noisy measurements on the target function. At each active learning iteration, the posterior distribution is first estimated using the current set of observations. A new sensor is then placed at the location corresponding to the maximum predicted uncertainty, and the resulting measurement is incorporated into the observation set for the subsequent inference step.

\begin{figure}[H]
  \centering
  \includegraphics[width=1\textwidth]{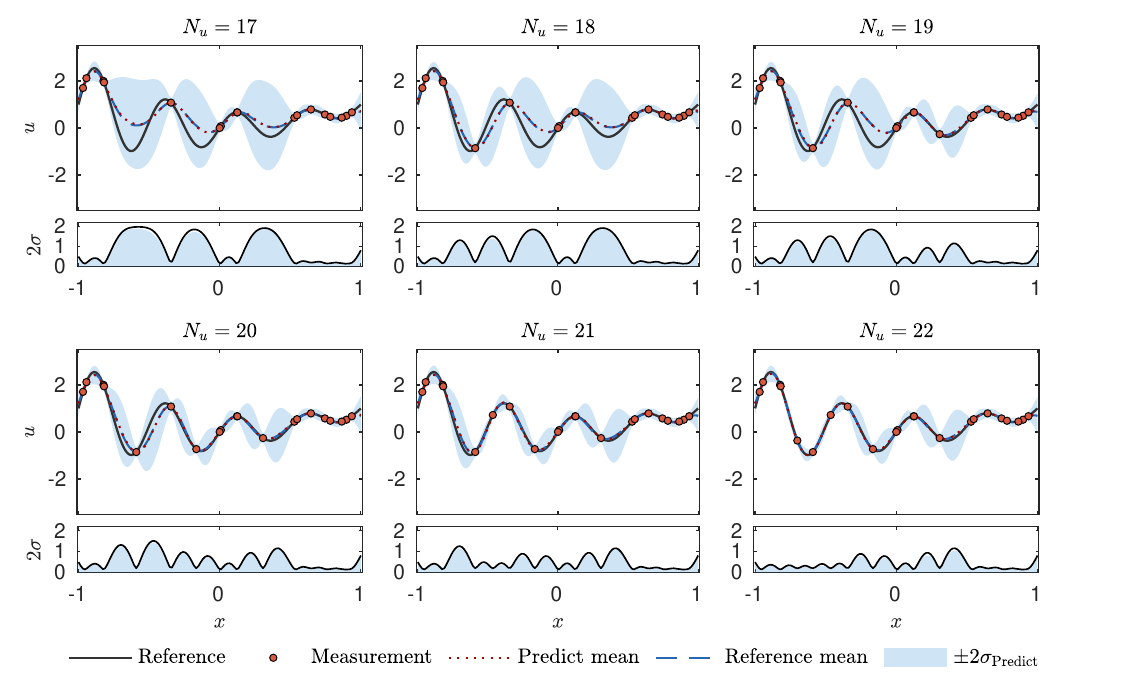}
  \caption{Flow-ABI for 1D regression problem: Active learning with the Gaussian process prior. }
  \label{fig:active_learning_progress}
\end{figure}

Fig. \ref{fig:active_learning_progress} illustrates the evolution of the posterior predictions throughout the active learning process. As additional sensors are sequentially deployed, the posterior mean progressively approaches the exact solution, while the predictive uncertainty is systematically reduced across the domain. In particular, regions associated with large posterior uncertainty are preferentially sampled, resulting in a progressively more informative observation set and improved reconstruction accuracy. After several active learning iterations, the posterior distribution becomes increasingly concentrated around the exact solution, indicating that the uncertainty estimates provided by the proposed framework effectively identify regions where additional measurements are most beneficial.

Notably, the active learning process requires repeated posterior updates as new observations become available. In conventional Bayesian inference frameworks, this typically entails solving a new inference problem after each sensor placement, which can be computationally expensive. In contrast, Flow-ABI performs posterior sampling directly through the pretrained Set-Conditioned Functional Posterior Model (SFPM), enabling rapid posterior updates without retraining or iterative sampling. Consequently, the proposed framework provides an efficient foundation for active learning, adaptive sensing, and optimal experimental design in scientific machine learning applications where real-time uncertainty quantification is essential.

\subsection{2D Function Approximation}
\label{sec:2d_regression}
We proceed to test the Flow-ABI on two-dimensional regression problems. The prior in this specific problem is assumed to be a zero-mean Gaussian process with an isotropic squared-exponential kernel for $u$, i.e. $u \sim \mathcal{GP}(0, \kappa(\bm{x}, \bm{x}'))$, which is expressed as:
\begin{equation}
  \kappa(\bm{x}, \bm{x}') = \exp\!\left(-\frac{(x_1 - x_1')^2}{2l_x^2} - \frac{(x_2 - x_2')^2}{2l_y^2}\right),~ \bm{x} \in [-1, 1]^2,
\end{equation}
with length scales $l_x = l_y = 0.2$. We aim to quantify the uncertainties arsing from noisy and gappy data in a two-dimensional regression case given this prior.

Following the training procedure described in Sec. \ref{sec:method}, we first train the functional prior model using samples drawn from the prescribed Gaussian process prior and subsequently train the Set-Conditioned Functional Posterior Model (SFPM) using paired samples generated from the learned prior model.  In particular, a training dataset consisting of $10,000$ independent realizations is generated from the specified Gaussian process prior, with each realization discretized on a $64\times64$ uniform grid over $\Omega$. These samples are used to train the function representation module (FRM): the coefficient network $\mathcal{NN}_C$ takes the field values on the $64\times64$ grid as input, while the basis network $\mathcal{NN}_B$ takes the continuous spatial coordinates $\bm{x}\in\Omega$ as input. Using the latent coefficient representations extracted by the trained FE, a one-step flow matching model is subsequently trained to approximate the prior distribution in the latent space, thereby enabling efficient generation of new prior realizations in the function space. Further, the learned prior model is used to generate synthetic observation–latent pairs for training the SFPM. At inference time, given a sparse observation set $\mathcal{D}_{\mathrm{obs}}$, the trained SFPM directly produces samples from the approximate posterior distribution, from which posterior statistics are estimated.

\begin{figure}[htbp]
  \centering
  \includegraphics[width=\textwidth]{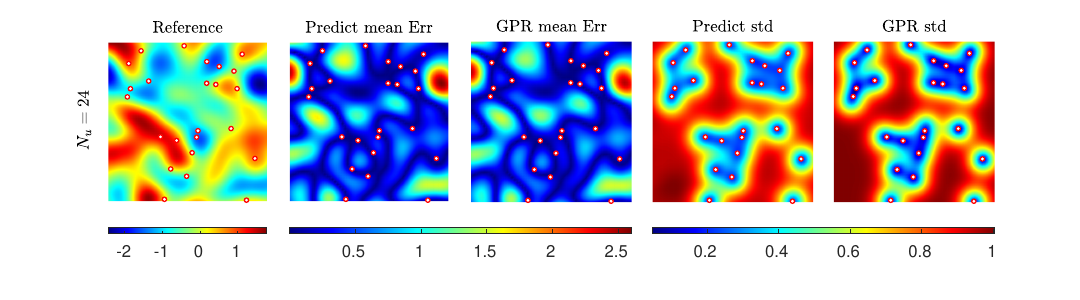}
  \includegraphics[width=\textwidth]{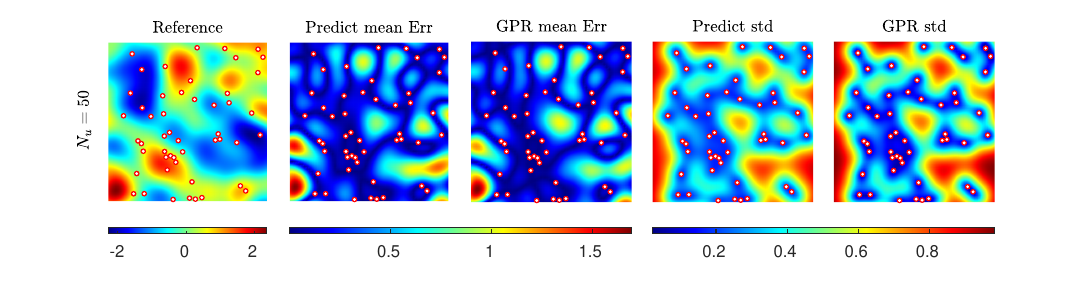}
  \includegraphics[width=\textwidth]{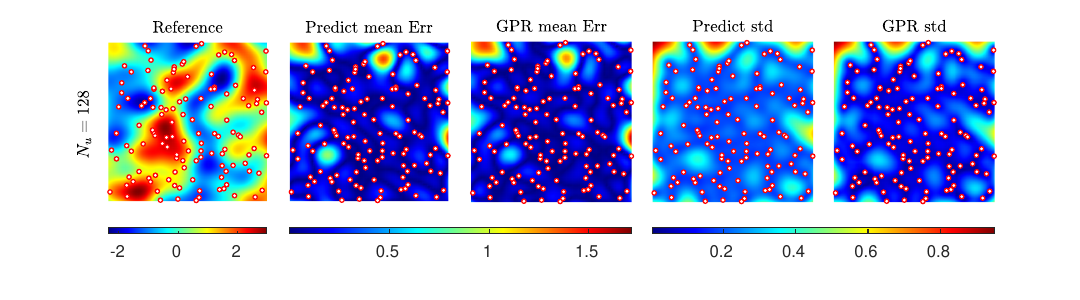}
  \includegraphics[width=\textwidth]{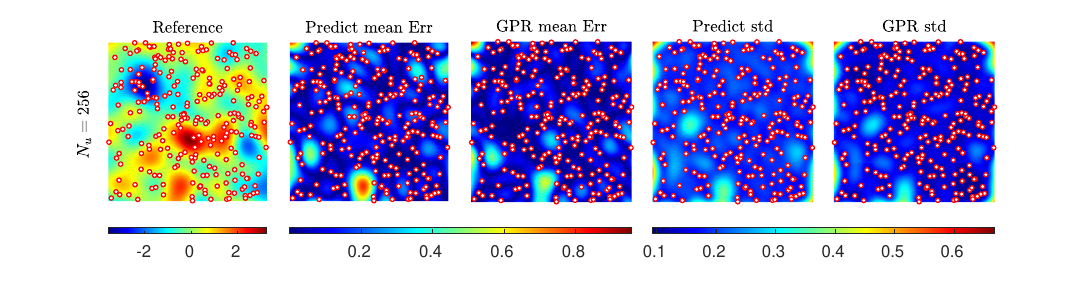}
  \caption{Flow-ABI for 2D regression problem:  From left to right in each row: the reference solution, the pointwise error between the predicted mean and the reference solution for Flow-ABI and GPR, and the predicted one standard deviation from the Flow-ABI and GPR.}
  \label{fig:2D_2}
\end{figure}

After the training, we employ four different cases to evaluate the accuracy of the Flow-ABI. In addition, the numbers as well as the locations of the measurements differ in each case. Also, the measurement noise is assumed to be Gaussian, i.e., $\mathcal{N}(0, 0.1^2)$. Since both the prior and the measurement noise are Gaussian, we therefore utilize results from the GPR as reference solutions as in case (a) of Sec. \ref{sec:regression}. As illustrated in Fig. \ref{fig:2D_2}, both the predicted mean and uncertainties from Flow-ABI closely match the reference solutions in all test cases, demonstrating the effectiveness of the Flow-ABI in two-dimensional regression problems.

\subsection{Nonlinear Elliptic Problem}
\label{sec:1d_pde}

We now consider to solve an inverse problem by integrating the Flow-ABI and the PINNs. Specifically, the governing equation tested here is a one-dimensional nonlinear elliptic equation
\begin{equation}
-\partial_x^2 u + \sin\bigl(u(x)\bigr)
= k f(x),
\qquad x\in[-1,1],
\label{eq:sine_gordon}
\end{equation}
where $k=0.005$ is a scaling parameter controlling the magnitude of the forcing term. The objective is to infer both the solution field $u(x)$ and the source term $f(x)$, together with their associated uncertainties, from sparse and noisy observations while satisfying the governing physical constraint.

Following Section~\ref{sec:regression}, we impose a zero-mean Gaussian process prior
$
u \sim \mathcal{GP} \left(0,\kappa(\bm{x},\bm{x}')\right)$
with the squared-exponential kernel
\[
\kappa(x,x') = 
\exp\left(
-\frac{|x-x'|^2}{2l^2},
\right), ~x\in[-1, 1]
\]
where the length scale is taken as $l=0.1$. Given realizations of $u$, the corresponding source term $f$ is obtained from Eq.~\eqref{eq:sine_gordon} using automatic differentiation, following the strategy adopted in \cite{raissi2019physics,yang2021b,zhou2025scalable,meng2022learning}. Consequently, a joint prior distribution over $(u,f)$ can be constructed.

The training of Flow-ABI follows the procedure described in Section~\ref{sec:method}. For the present example, the functional prior of $u$ is trained using $10,000$ realizations sampled from the prescribed Gaussian process prior, each discretized at 128 uniformly distributed locations over $[-1,1]$. The learned prior accurately reproduces the covariance structure of the target Gaussian process, similar as the result in Sec. \ref{sec:regression}, and the corresponding validation results are omitted here for brevity. Further, Given realizations of $u$, the corresponding source term $f$ is obtained from Eq.~\eqref{eq:sine_gordon} using AD, as aforementioned. Consequently, a joint prior distribution over $(u,f)$ can be constructed. The trained prior is subsequently employed to generate paired samples for training the Set-Conditioned Functional Posterior Model (SFPM). Unless otherwise stated, additive measurement noise is assumed to follow independent Gaussian distributions with standard deviations $\sigma_u=\sigma_f=0.1$.

During inference, the SFPM receives a sparse observation set $\mathcal{D}_{\mathrm{obs}}$ consisting of noisy measurements of both $u$ and $f$, and directly generates posterior samples of the latent variables, which are subsequently decoded into joint posterior realizations of $(u,f)$ over the entire domain. Also, the trained model naturally accommodates varying numbers and spatial locations of sensors without retraining.

\begin{figure}[H]
  \centering
  \includegraphics[width=\textwidth]{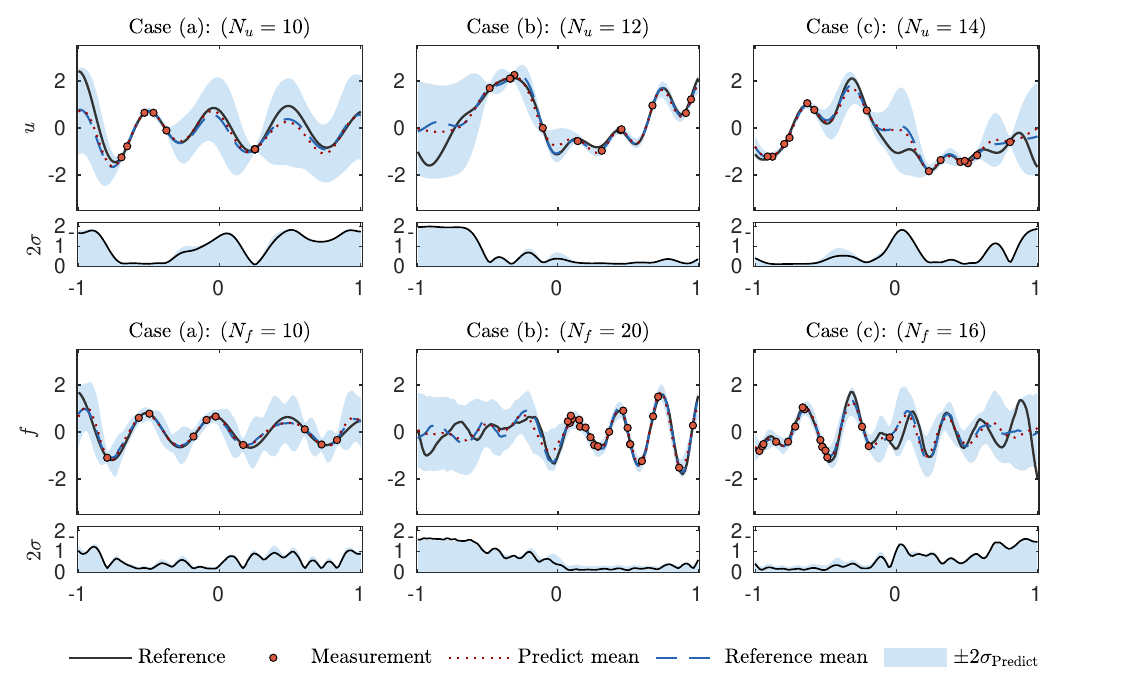}
  \caption{Flow-ABI for nonlinear elliptic problem: The two plots in each column correspond to the reconstructed $u$ and the associated $f$ with uncertainties, respectively.}
  \label{fig:pde_uf}
\end{figure}

To evaluate the accuracy of Flow-ABI, we consider three representative test cases. In Case~(a), the target solution is prescribed as
\begin{equation*}
    u(x)
    =
    2e^{-(x+1)}\sin(4\pi x)
    -
    0.5\sin(3\pi x + x^2),
\end{equation*}
and reconstructed from 16 noisy measurements of both $u$ and $f$. In Cases~(b) and~(c), the target functions are randomly sampled from the same Gaussian process prior with length scale $l=0.1$ and measurement noise standard deviation $\sigma=0.1$. The three cases differ in the number and spatial distribution of the sensors, thereby providing a more challenging assessment of the model's ability to generalize across different observation configurations. For each test case, $10,000$ posterior samples are generated using both Flow-ABI and the reference HMC method, from which the posterior mean and standard deviation are estimated. As shown in Fig.~\ref{fig:pde_uf}, the predictions obtained by Flow-ABI are in excellent agreement with the HMC reference for both the posterior mean and uncertainty estimates across all three cases. In particular, the proposed method accurately captures the non-Gaussian posterior distributions induced by the nonlinear term $\sin(u)$ while maintaining well-calibrated uncertainty quantification. Moreover, consistent reconstruction of both $u$ and $f$ is achieved despite substantial variations in sensor numbers and locations, demonstrating the robustness and generalization capability of Flow-ABI for coupled field inference under varying observation configurations.


In Table~\ref{tab:time_comparison}, we present the wall-clock time required to generate $10,000$ posterior samples using Flow-ABI and HMC. The proposed approach achieves an approximately $199 \times$ speed-up compared with single-chain HMC and a $112 \times$ speed-up compared with four-chain parallel HMC, while maintaining comparable posterior accuracy. This computational advantage is particularly pronounced for the nonlinear elliptic inverse problem considered here, where each HMC iteration requires repeated evaluations of the nonlinear forward model and its associated gradients. In contrast, once trained, Flow-ABI directly generates posterior samples through a single forward pass of the learned generative model, eliminating the need for HMC iterations during inference. As a result, posterior distributions can be estimated within seconds, enabling rapid uncertainty quantification and efficient assimilation of newly acquired observations.

\begin{table}[H]
  \centering
  \caption{Flow-ABI for nonlinear elliptic problem: Computational time (in seconds) and speed-up of the proposed model versus HMC when drawing $10,000$ posterior samples.}
  \label{tab:time_comparison}
  \begin{tabular}{l|c|cc|cc}
    \toprule
    & Flow-ABI & \makecell{NUTS \\ (single chain)} & Speed-up & \makecell{NUTS \\ (4 chains)} & Speed-up \\
    \midrule
    Walltime (s) & 0.69 & 137.39 & 199.11 & 77.86 & 112.84 \\
    \bottomrule
  \end{tabular}
\end{table}



\subsection{Neural Operator: 2D Darcy Flow in Heterogeneous Porous Media}
\label{sec:darcy}
We further test the Flow-ABI using an example of flow through heterogeneous porous media, which is a widely used benchmark for inverse problem. The specific problem consider here is governed by the following two-dimensional equation \cite{zheng2020physics}:
\begin{equation}
  -\nabla \cdot \bigl(K(\bm{x})\,\nabla h(\bm{x})\bigr) = f(\bm{x}), \quad \bm{x} = (x, y) \in [-1,1]^2,
  \label{eq:Darcy}
\end{equation}
which subjects to the boundary conditions
\begin{equation}
  h(x,-1) = 1, \quad h(x,1) = 0, \quad \partial_{\bm{n}}h(-1,y) = 0, \quad \partial_{\bm{n}}h(1,y) = 0,
\end{equation}
where $K(\bm{x})$ denotes the hydraulic conductivity, $h(\bm{x})$ the hydraulic head, and $f(\bm{x}) = 1$ a uniform source term. To capture the spatial variability characteristic of realistic conductivity fields, $K(\bm{x})$ is modeled following \cite{zheng2020physics} as:
\begin{equation}
  K(\bm{x}) = \exp\bigl(F(\bm{x})\bigr),
\end{equation}
where $F(\bm{x})$ represents the realization of the following prior, i.e., $F \sim \mathcal{GP}(0, \kappa(\bm{x}, \bm{x}'))$, with the covariance specified as an anisotropic squared-exponential kernel:
\begin{equation}
  \kappa(\bm{x}, \bm{x}') = \exp\!\left(-\frac{(x-x')^2}{2l_x^2} - \frac{(y-y')^2}{2l_y^2}\right), \quad \bm{x}, \bm{x}' \in [-1,1]^2,
\end{equation}
in which $l_x = 0.07$ and $l_y = 0.2$. For this specific problem, we assume that we have noisy measurements on $K$ and $h$, the objective is to reconstruct the full fields of conductivity $K(\bm{x})$ and hydraulic head $h(\bm{x})$ and their associated uncertainties from solely these sparse observations while constrained the underlying physics of the governing flow equations. Generally, $F(\bm{x})$ is employed in the computations rather than $K$ to guarantee that the obtained $K$ is nonnegative \cite{zheng2020physics,meng2022learning,psaros2023uncertainty,zou2024neuraluq}.

In order to showcase the flexibility as well as effectiveness of Flow-ABI, we will integrate the Flow-ABI with the neural operator, more specifically the Fourier neural operator (FNO) \cite{li2020fourier}, to solve the above inverse problem. Specifically, we randomly draw samples for $F(\bm{x})$ from the corresponding prior, and then we employ the PDEToolboox in Matlab to solve the governing equation to obtain the paired data ${F(\bm{x})_i, h_i(\bm{x})}^N_{i=1}$. These data are  subsequently utilized to train the FNO to learn the mapping from $F(\bm{x})$ to $h(\bm{x})$, i.e., $h(\bm{x}) = \mathcal{N}(F(\bm{x}))$.


With the pretrained FNO, we then integrate the Flow-ABI with it for solving inverse Darcy problems. The training of Flow-ABI proceeds according to the two-stage scheme detailed in Sec. \ref{sec:framework}. In the first stage, the functional prior is trained using $N=10,000$ realizations for $F(\bm{x})$ drawn from the prescribed Gaussian process prior and discretized on a $64\times64$ uniform grid. With the well trained prior for $F(\bm{x})$, we are able to obtain samples for $F(\bm{x})$, and thus the prior for $h(\bm{x})$ can be obtained based on the pretrained FNO, i.e., $h(\bm{x}) = \mathcal{N}(F(\bm{x}))$. Subsequently, The trained prior is used to generate paired samples for training the Set-Conditioned Functional Posterior Model (SFPM). In this case, the measurements of both $F$ and $h$ are corrupted by additive Gaussian noise with a standard deviation of $\sigma = 0.05$. During inference, the SFPM takes a sparse observation set $\mathcal{D}_{\mathrm{obs}}$ which consisting of noisy measurements of both $F$ and $h$ as input, and produces posterior samples of the latent variables. These are then decoded into posterior realizations of log-conductivity field $F$ over the entire domain and get the hydraulic head $h$.

To evaluate the accuracy of Flow-ABI, we consider three representative test cases with varying numbers and spatial configurations of sensors. Since the posterior distribution does not admit a closed-form expression due to the nonlinear forward map, posterior samples obtained via HMC serve as the reference solution. For each case, $10,000$ posterior samples are drawn using both Flow-ABI and HMC, from which the posterior mean and standard deviation are estimated.

\begin{figure}[H]
  \centering
  \renewcommand{\thesubfigure}{}
  \subfigure[]{\label{fig:darcy_A}
  \includegraphics[width=0.82\textwidth]{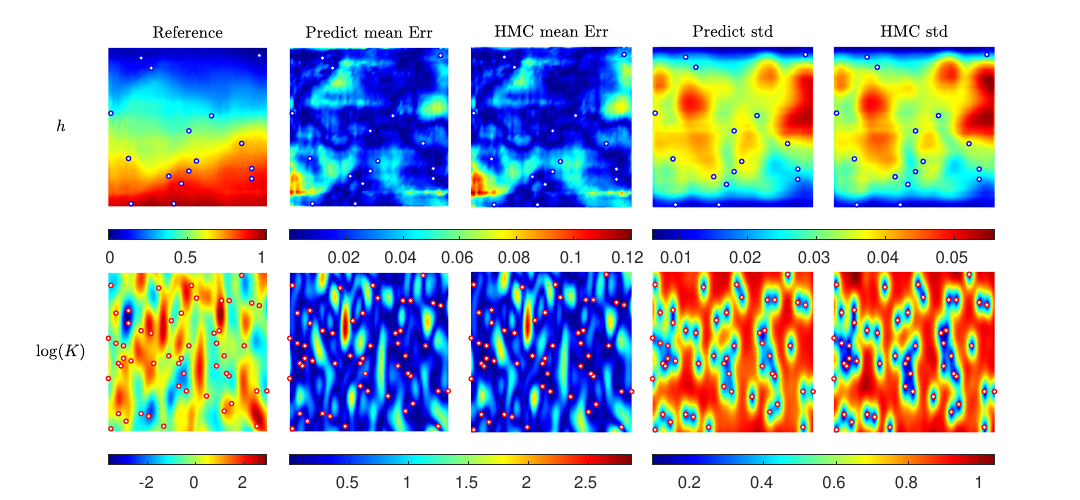}}
  \subfigure[]{\label{fig:darcy_B}
  \includegraphics[width=0.82\textwidth]{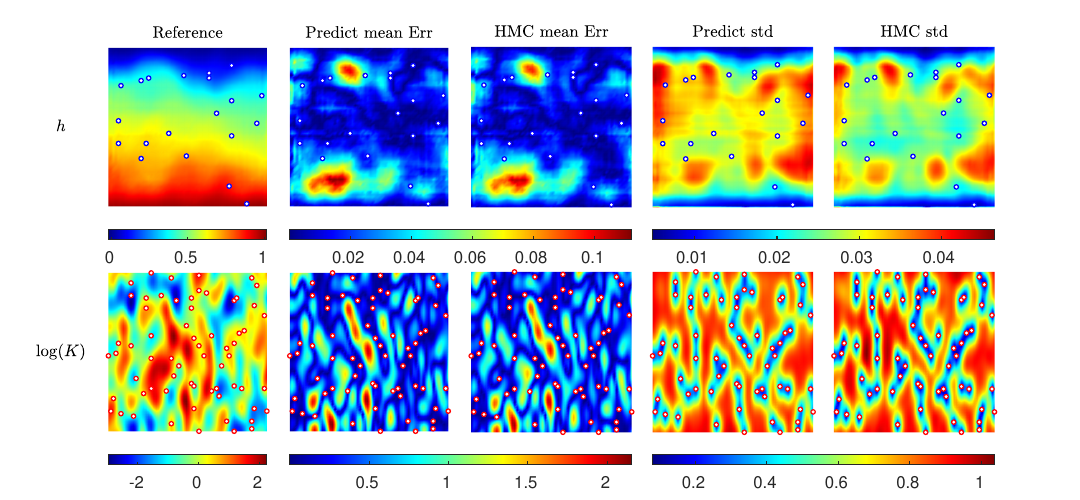}}
  \subfigure[]{\label{fig:darcy_C}
  \includegraphics[width=0.82\textwidth]{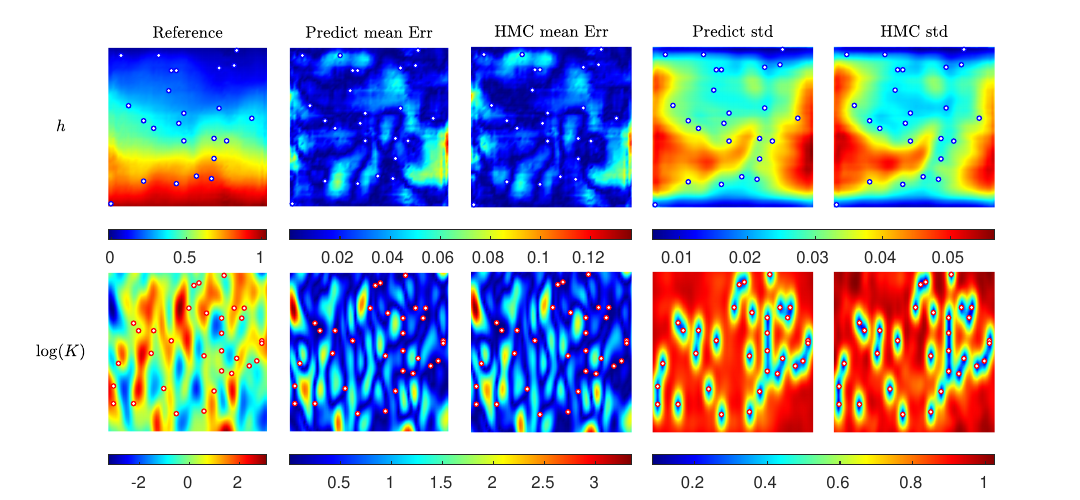}}
  \caption{Flow-ABI for 2D Darcy problem: Predicted mean and uncertainties for the log-conductivity field $F$ and the hydraulic head $h$. Observations of both $F$ and $h$ are corrupted by additive Gaussian noise with standard deviation $\sigma = 0.05$.}
  \label{fig:Darcy2}
\end{figure}

As illustrated in Fig.~\ref{fig:Darcy2}, Flow-ABI closely matches the HMC reference in both posterior mean estimation and uncertainty quantification across all three sensor configurations. In particular, the proposed method accurately recovers the anisotropic spatial structure of the log-conductivity field, with elongated correlation patterns along the $y$-direction faithfully reproduced in the posterior mean. The predicted standard deviation is also well-calibrated, with regions of higher uncertainty correctly identified in areas far from sensors or subject to greater prior variability. These results demonstrate that the framework scales effectively to 2D inverse problems involving non-Gaussian posteriors and implicit forward operators approximated by neural surrogates.

Table~\ref{tab:time_comparison_darcy} reports the wall-clock time required to draw $10,000$ posterior samples using Flow-ABI and HMC, respectively. The proposed approach achieves an approximately $1039 \times$ speed-up compared with single-chain HMC and a $393 \times$ speed-up compared with four-chain parallel HMC, while maintaining comparable posterior accuracy, reducing the sampling time from over 7.6 minutes to under 1.16 seconds with the four-chain parallel HMC. This efficiency gain is particularly significant for applications requiring repeated posterior evaluations, such as optimal sensor placement or real-time subsurface monitoring, where the amortized nature of Flow-ABI eliminates the need for retraining or iterative sampling under each new observation configuration.

\begin{table}[H]
  \centering
  \caption{Flow-ABI for 2D Darcy problem: Computational time (in seconds) and speed-up of the Flow-ABI versus HMC when drawing $10,000$ posterior samples.}
  \label{tab:time_comparison_darcy}
  \begin{tabular}{l|c|cc|cc}
    \toprule
    & Flow-ABI & \makecell{NUTS \\ (single chain)} & Speed-up & \makecell{NUTS \\ (4 chains)} & Speed-up \\
    \midrule
    Walltime (s) & 1.16 & 1205.65 & 1039.35 & 456.14 & 393.10 \\
    \bottomrule
  \end{tabular}
\end{table}

\section{Summary}
\label{sec:summary}
In this work, we proposed Flow-ABI, a conditional flow-matching framework for amortized Bayesian inference in regression and inverse PDE problems. Unlike conventional Bayesian approaches that require solving a new inference problem for every observation set, Flow-ABI learns a set-to-distribution mapping during an offline training stage, enabling near-real-time posterior sampling for previously unseen observations without retraining or iterative optimization. By performing uncertainty quantification directly in function spaces, the proposed framework is applicable to a broad range of scientific machine learning tasks involving functional unknowns.

A key feature of Flow-ABI is its set-conditioned posterior representation, which naturally accommodates varying numbers, locations, and types of measurements. Consequently, the learned posterior operator is both permutation-invariant and discretization-invariant, allowing the framework to generalize across different sensor configurations and observation discretizations. Furthermore, Flow-ABI can be seamlessly integrated with both physics-informed neural networks and neural operators, providing a general and flexible approach to uncertainty-aware inverse modeling in PDE-governed systems.

Through extensive numerical experiments on representative regression and inverse PDE benchmarks, we demonstrated that Flow-ABI accurately captures complex posterior distributions, including highly non-Gaussian uncertainties, while achieving more than two orders of magnitude speedup compared with the conventional gold-standard Bayesian inference method, namely Hamiltonian Monte Carlo (HMC). These results establish the Flow-ABI as an effective, scalable, and computationally efficient framework for uncertainty quantification in scientific machine learning.

Several directions warrant further investigation. Future work will focus on extending the proposed framework to high-dimensional spatio-temporal systems, multi-physics and multi-fidelity settings, and sequential data assimilation problems. Another promising direction is the integration of adaptive sensing and experimental design strategies, where the learned set-conditioned functional posterior model can be leveraged to guide data acquisition and sensor placement in real time. More broadly, we envision Flow-ABI as a foundational building block for uncertainty-aware scientific machine learning systems, enabling rapid, reliable, and scalable decision-making in complex physical environments.

\section*{Acknowledgements}
S. Z. and X. M. acknowledge the support of the National Natural Science Foundation of China (No. 12201229) and the Interdisciplinary Research Program of HUST (No. 2024JCYJ003). X. M. also acknowledges the support of the Xiaomi Young Talents Program.  L. G. acknowledges the support of the National Natural Science Foundation of China (No. 92270115, 12071301).

\appendix

\section{Details on the computations}
\label{sec:computations}
In all the cases discussed in the Results section, the AdamW optimizer and a cosine annealing schedule for the learning rate are utilized for training all neural networks.  Additional details regarding the architectures and training steps for functional prior and posterior models are provided in Tables  \ref{table:deeponet_arch} - \ref{tab-Trans_batch}.  Furthermore, during the distillation of the prior, the flow matching process is solved using a 20-step fourth-order Runge-Kutta (RK4) method. For the posterior, a 5-step RK4 method is employed. We observed that 5 steps are sufficient to generate high-quality posteriors, and further increasing the number of steps yields marginal improvements. 
In Section \ref{sec:regression}, we investigated the error associated with permuted observations. We observed that as the number of RKH steps increases, the error steadily decreases, yielding $5.3 \times 10^{-5}$ at 5 steps, $3.7 \times 10^{-5}$ at 10 steps, and $1.8 \times 10^{-5}$ at 40 steps.

In addition, the details of Fourier Neural Operator (FNO) using in the Sec. \ref{sec:darcy} are as follows: the employed FNO consists of a lifting layer in the $64\times64$ uniform grid, four consecutive Fourier layers, and a projection layer. Specifically, the input features are first lifted to a latent space with a channel width of 12. Across the four Fourier layers, the network operates in the frequency domain by retaining only the maximal modes of 32 and 12 along the first and second spatial dimensions, respectively. The ReLU activation function is applied after the residual connections in the first three Fourier layers and within the projection block. Finally, the projection layer utilizes a two-layer perceptron with a hidden dimension of 128 to map the latent representations to the desired single-channel output.

\begin{table}[H]
\caption{
Architecture and training steps of functional encoder in each case. Specifically, both $\mathcal{NN}_{B}$ and $\mathcal{NN}_C$ are fully-connected neural networks (FNN).}
\centering
\footnotesize
\renewcommand{\arraystretch}{1.2}
\setlength{\tabcolsep}{3pt} 
\begin{tabular}{l l l l l l}
\toprule
  & \multicolumn{2}{l}{$\mathcal{NN}_C$}  & \multicolumn{2}{l}{$\mathcal{NN}_B$} & \multirow{2}{*}{\thead{Training\\steps}} \\
  \cmidrule(lr){2-3} \cmidrule(lr){4-5}
  & width $\times$ depth  & Activation & width $\times$ depth & Activation \\
\midrule
  Sec. \ref{sec:regression} & $64 \times 5$   & ReLU & $64 \times 5$ & tanh &  200,000 \\
  Sec. \ref{sec:2d_regression} &  $256 \times 5$  & ReLU & $256 \times 5$ & tanh &  200,000 \\
  Sec. \ref{sec:1d_pde} & $64 \times 5$ & ReLU & $64 \times 5$ & tanh & 200,000  \\
  Sec. \ref{sec:darcy} & $256 \times 5$ & ReLU & $256 \times 5$ & tanh & 200,000  \\
\bottomrule
\end{tabular}
\label{table:deeponet_arch}
\end{table}

\begin{table}[H]
\caption{Batch size and training points for each case in functional encoder.}
\centering
\footnotesize
\renewcommand{\arraystretch}{1.2}
\setlength{\tabcolsep}{5.5pt} 
\begin{tabular}{l l l l}
\toprule
  & Batch size & Training points for each snapshot &  learning rate \\
\midrule
  Sec. \ref{sec:regression} & 1280 & 2048 (equidistantly distributed in $[-1,1]$) & $1 \times 10^{-3}$\\
  Sec. \ref{sec:2d_regression} & 400 & 2048 (uniformly distributed in $[-1,1]^2$) & $1 \times 10^{-3}$\\
  Sec. \ref{sec:1d_pde} & 1280 & 2048 (equidistantly distributed in $[-1,1]$) & $1 \times 10^{-3}$\\
  Sec. \ref{sec:darcy} & 400 & 2048 (uniformly distributed in $[-1,1]^2$) & $1 \times 10^{-3}$\\
\bottomrule
\end{tabular}
\end{table}

\begin{table}[H]
\caption{Architectures of based flow-matching generative models in the FPM, which are all FNNs.}
\centering
\footnotesize
\renewcommand{\arraystretch}{1.2}
\setlength{\tabcolsep}{7.0pt} 
\begin{tabular}{l c c c}
\toprule
& width $\times$ depth  &  FFN Activation  \\
\midrule
Sec. \ref{sec:regression} & $64 \times 5$  & GeLU \\
Sec. \ref{sec:2d_regression} & $256 \times 5$  & GeLU \\
Sec. \ref{sec:1d_pde} & $64 \times 5$ &  GeLU \\
Sec. \ref{sec:darcy} & $256 \times 5$ &  GeLU \\
\bottomrule
\end{tabular}
\end{table}

\begin{table}[H]
\caption{Batch size, training steps and learning rate (lr) of based flow-matching in the FPM.}
\centering
\footnotesize
\renewcommand{\arraystretch}{1.2}
\setlength{\tabcolsep}{5.5pt} 
\begin{tabular}{l c c c}
\toprule
  & Batch size &  learning rate & Training steps \\
\midrule
  Sec. \ref{sec:regression} & 1280  & $1 \times 10^{-3}$ & 200,000\\
  Sec. \ref{sec:2d_regression} & 400  & $1 \times 10^{-3}$ & 200,000\\
  Sec. \ref{sec:1d_pde} & 1280  & $1 \times 10^{-3}$ & 200,000\\
  Sec. \ref{sec:darcy} & 400  & $1 \times 10^{-3}$ & 200,000\\
\bottomrule
\end{tabular}
\end{table}

\begin{table}[H]
\caption{Architectures of the one-step generators in FPM, which are all FNNs.}
\centering
\footnotesize
\renewcommand{\arraystretch}{1.2}
\setlength{\tabcolsep}{7.0pt} 
\begin{tabular}{l c c c }
\toprule
& width $\times$ depth & Activation  \\
\midrule
Sec. \ref{sec:regression} & $64 \times 5$  & GeLU  \\
Sec. \ref{sec:2d_regression} & $256 \times 5$  & GeLU \\
Sec. \ref{sec:1d_pde} & $64 \times 5$ &  GeLU \\
Sec. \ref{sec:darcy} & $256 \times 5$ &  GeLU\\
\bottomrule
\end{tabular}
\end{table}

\begin{table}[H]
\caption{Batch size, training steps and learning rate (lr) of the one-step generator in each test case of the FPM.}
\centering
\footnotesize
\renewcommand{\arraystretch}{1.2}
\setlength{\tabcolsep}{5.5pt} 
\begin{tabular}{l c c c}
\toprule
  & Batch size & Training steps&  learning rate \\
\midrule
  Sec. \ref{sec:regression} & 1280 & 200,000 & $1 \times 10^{-3}$\\
  Sec. \ref{sec:2d_regression} & 400 & 200,000 & $1 \times 10^{-3}$\\
  Sec. \ref{sec:1d_pde} & 1280 & 200,000 & $1 \times 10^{-3}$\\
  Sec. \ref{sec:darcy} & 400 & 200,000 & $1 \times 10^{-3}$\\
\bottomrule
\end{tabular}
\end{table}

\begin{table}[H]
\caption{Architecture of the diffusion Transformer in the SFPM of each test case.}
\centering
\footnotesize
\renewcommand{\arraystretch}{1.2}
\setlength{\tabcolsep}{7.0pt} 
\begin{tabular}{l c c c c c}
\toprule
& Model-dim & Block Num. & Heads & FFN hidden dim & FFN Activation\\
\midrule
Sec. \ref{sec:regression} & 256 & 4 & 8 & 1024 & GeLU\\
Sec. \ref{sec:2d_regression} & 256 & 6 & 8 & 1024 & GeLU\\
Sec. \ref{sec:1d_pde} & 256 & 4 & 8 & 1024 & GeLU\\
Sec. \ref{sec:darcy} & 256 & 8 & 8 & 1024 & GeLU\\
\bottomrule
\end{tabular}
\end{table}

\begin{table}[H]
\caption{Batch size and training steps for the SFPM}
\centering
\footnotesize
\renewcommand{\arraystretch}{1.2}
\setlength{\tabcolsep}{5.5pt} 
\begin{tabular}{l c c c}
\toprule
  & Batch size & Training steps&  learning rate \\
\midrule
  Sec. \ref{sec:regression} & 1024 & 200,000 & $1 \times 10^{-3}$\\
  Sec. \ref{sec:2d_regression} & 1024 & 200,000 & $1 \times 10^{-3}$\\
  Sec. \ref{sec:1d_pde} & 1024 & 200,000 & $1 \times 10^{-3}$\\
  Sec. \ref{sec:darcy} & 1024 & 200,000 & $1 \times 10^{-3}$\\
\bottomrule
\end{tabular}
\end{table}

\begin{table}[H]
\caption{Maximum number of sensors for the SFPM in each test case.}
\label{tab-Trans_batch}
\centering
\footnotesize
\renewcommand{\arraystretch}{1.2}
\setlength{\tabcolsep}{5.5pt} 
\begin{tabular}{l c c c c}
\toprule
 & Sec. \ref{sec:regression} & Sec. \ref{sec:2d_regression} & Sec. \ref{sec:1d_pde} & Sec. \ref{sec:darcy} \\
\midrule
Maximum number of sensors & $N_u=32$ & $N_u=256$ & $N_u=32,N_f=32$ & $N_h=64,N_F=64$ \\
\bottomrule
\end{tabular}
\end{table}

\bibliographystyle{unsrt}
\bibliography{refs}





\end{document}